\documentclass[12pt]{article}

\usepackage{a4wide}
\usepackage{amsmath}
\usepackage{amsthm}
\usepackage{amsfonts}
\usepackage{amssymb}
\usepackage{bbm}
\usepackage{cite}
\usepackage{graphicx}

\newcommand{\N}{\mathbbm{N}}

\newcommand{\R}{\mathbbm{R}}

\newcommand{\F}{\mathcal{F}}
\newcommand{\D}{\Delta}
\renewcommand{\l}{\lambda}
\renewcommand{\L}{\Lambda}
\newcommand{\eg}{e.g.\ }
\newcommand{\ie}{i.e.\ }
\newcommand{\e}{\mathrm{e}}
\newcommand{\ab}{(\textit{a}) and (\textit{b})}
\newcommand{\ii}{\mathrm{i}}
\newcommand{\order}[1]{{O({#1})}}

\newtheorem{theorem}{Theorem}[section]
\newtheorem{lemma}[theorem]{Lemma}
\newtheorem{remark}[theorem]{Remark}

\DeclareMathOperator{\Tr}{Tr}

\numberwithin{equation}{section}

\begin{document}
\thispagestyle{empty}
\vspace*{-80pt} 
{ \hspace*{\fill} Preprint-KUL-TF-2002/09} 
\vspace{55pt} 
\begin{center} 
{\Large   Bose-Einstein Condensation for Homogeneous 
\\[6pt]
 Interacting Systems with a One-Particle Spectral Gap 
  }
 \\[30pt]  
 
{\large  
    J.~Lauwers,
       \footnote{Email: {\tt joris.lauwers@fys.kuleuven.ac.be}}
    A.~Verbeure,
       \footnote{Email: {\tt andre.verbeure@fys.kuleuven.ac.be}}
    and V.~A.~Zagrebnov 
       \footnote{Email: {\tt zagrebnov@cpt.univ-mrs.fr}}}
       \footnotetext{on leave of absence from: Universit\'e de la 
       M\'editerran\'ee and Centre de Physique Th\'eorique, 
       CNRS-Luminy-Case 907, 13288 Marseille, Cedex 09, France}
\\[30pt]   
Instituut voor Theoretische Fysica,  
Katholieke Universiteit Leuven,  
Celestijnenlaan~200D,  
B-3001 Leuven, Belgium\\[25pt]
{July 24, 2002, modified December 10, 2002 }\\[25pt]
\end{center} 
\begin{abstract}\noindent
We prove rigorously the occurrence of zero-mode Bose-Einstein condensation for
a class of continuous homogeneous systems of boson particles with superstable
interactions.  This is the first example of a translation invariant continuous
Bose-system, where the existence of the Bose-Einstein condensation is proved 
rigorously for the case of non-trivial two-body particle interactions, provided
there is a large enough one-particle excitations spectral gap. The idea of proof consists of comparing the system with 
specially tuned soluble models. 
\\[15pt]
\end{abstract}

\newpage

\section{Introduction}

Rigorous proofs of Bose-Einstein condensation (BEC) for realistic systems is an
ongoing challenge for almost eight decades with renewed interest after the 
successful experiments with trapped gases of alkali metals.
In particular establishing a proof of condensation for a system of interacting
particles turns out to be a real hard problem. However it is believed that this
the real issue, in view of its experimental realisation in superfluid 
${}^{4}\mathrm{He}$. 

For this work, we were 
inspired by the recent results on condensation for
trapped gases, \ie inhomogeneous systems because of the external trapping
fields \cite{lieb:2002} and by the result on Bose-Einstein condensation for 
systems with a gap in the one-particle excitation spectrum with a Van der Waals 
family of two-body potentials by Buffet, de Smedt and Pul\'e \cite{buffet:1983b}.  
Although these two exact results are \textit{a priori} unrelated, the first one
being for inhomogeneous systems and the second one for homogeneous systems, both
start from systems with a gap in the one-particle 
excitations spectrum. Heuristically, in both cases one might start from a 
system with a gap in the spectrum and with condensation in the groundstate.
Contrary to systems without a gap, one can anticipate that the condensation for systems with a gap is stable 
under small perturbations. One might understand that condensation in the 
groundstate, which is energetically isolated by a gap, can survive the 
switching-on of a gentle interaction, and that fluctuations must be of 
macroscopic size to overcome this gap and lift particles out of the isolated 
groundstate.

In this note, we turn our attention to homogeneous continuous systems (\ie
without external trapping fields) of interacting bosons in the standard 
thermodynamic limit. We prove that there is Bose-Einstein condensation for high
enough density and low enough temperature, provided that there is a large enough
gap in the one-particle excitations spectrum. To our knowledge, this is the 
first example of a proof of condensation for systems with general 
two-body potentials.

We remark that for the trapped systems one proves in \cite{lieb:2002} in fact
macroscopic occupation of the ground state, which is the main property of BEC.
However, as these systems are not homogeneous in any sense, there is no notion 
of phase transition present. The presence of a trap yields \emph{a fortiori} a
discrete one-particle spectrum. In our models we also assume a gap in the
one-particle spectrum, and we prove standard BEC including the phase transition
(gauge symmetry breaking, off diagonal long range order,\ldots) which follows
directly from the homogeneity of the system. Therefore 
we consider our result as a step forward to the realisation of a bridge  
between the homogeneous and the trapped case, leading to a concept of 
phase transition also for the trapped case. In particular it might link the 
notion of the standard thermodynamic limit and the so-called Gross-Pitaevskii 
limit which is used in the trap case.

Another challenging problem posed by our results is of course: can one close the
gap ? \ie can one prove that BEC persists if one takes the limit of the gap
tending to zero.

These were our starting ideas. First we define our systems under consideration.
Let us consider a gas of interacting bosons in hypercubic boxes $\L = 
[-L/2,L/2]^\nu \subset \R^\nu$ of dimension $\nu \geq 1$ 
with periodic boundary conditions. The generalisation to other shapes for $\L$ 
is however readily done.  Denote by $V= L^\nu$ the volume of the box $\L$. 
The Hamiltonian for the volume $\L$ of this system in the boson Fock-space 
$\F_B$ reads
\begin{equation} \label{H-int}
H^\D_{\L, g} = T^\D_\L -\mu N_\L + g U_\L, \qquad g > 0,
\end{equation} 
where $T^\D_\L$ is the kinetic energy operator with the gap $\Delta > 0$ in its
spectrum,
\begin{equation}\label{T-Delta}
T^\D_\L = \sum_{k \in \L^*} \varepsilon_k a^\dagger_k a_k -\Delta
a^\dagger_0 a_0, \qquad \varepsilon_k = \frac{\hbar^2k^2}{2m}.
\end{equation}
We take units $\hbar^2/2m =1$,
the sum $k$ runs over the set $\L^*$, dual to $\L$, \ie 
\begin{equation*}
\L^* = \left\{ k
\in \R^\nu; k_\alpha = 2\pi n_\alpha /L; n_\alpha = 0,\pm 1,\ldots; \alpha
=1,2,\ldots,\nu \right \}.
\end{equation*} 
The operators $a^\dagger_k$ and $a_k$ are the Bose
creation and annihilation operators for mode $k\in\L^*$, the number 
operators are denoted by $N_k = a^\dagger_ka_k $, and  the total number operator 
in the volume $\L$ by $N_\L = \sum_{k \in \L^*}N_k$.

We assume \textit{a priori} the presence of a gap $\D$ in the one-particle
excitations spectrum, isolating the lowest energy level.
This might seem rather artificial, however note that the presence 
of such a gap can be realised in several ways. A gap can be created using 
attractive boundary conditions \cite{robinson:1976, landau:1979, lauwers:2002}, 
\ie considering systems in containers $\L$ with \textit{sticky} boundary
conditions. But we have to notice that the Bose condensate is not homogeneous
in this case.
Another possibility is to assume that part of the particle interaction has 
caused this gap and is as such effectively incorporated \cite{zagrebnov:2001}. 
The idea of considering a gap is not new. In his book \cite{london:1954} 
F.~London attempted to introduce the gap on heuristic grounds to
clarify some of the spectral properties of superfluid ${}^{4}\mathrm{He}$. 

The interaction between the particles is modelled by the  
two-body interaction operator
\begin{equation}\label{U-Lambda}
U_\Lambda =  \frac{1}{2}\int_{\Lambda^2}\!\mathrm{d}x\mathrm{d}y\ a^\dagger (x)
a^\dagger (y) v( x - y) a(y)a(x),
\end{equation}
where 
$a^\dagger (x),a^\dagger (y)$ and $a(y),a(x)$ are the creation and annihilation 
operators for the Bose particles at $x,y \in \R^\nu$. 
Below we assume that the pair interaction potential $v(x)$ verifies
the following conditions:
\begin{enumerate}
\item[(\textit{a})]
$v:\R^\nu \to \R^1$ is a real, positive-type function from
$L^1(\R^\nu)$. As it is known (the Bochner theorem \cite{reed:1975}), a 
continuous function is of
positive type if and only if it is the Fourier transform of a positive measure
$\mathrm{d}\mu$ of finite total mass on $\R^\nu$:
\begin{equation*}
v(x) = \int_{\R^\nu}\! \mathrm{d}\mu(q)\ \e^{\ii q x} = v(-x).
\end{equation*} 
\item[(\textit{b})]
Since $v \in L^1(\R^\nu)$, the Fourier transform $\hat{v}(q)$ exists, and we
suppose that
 \begin{equation}\label{vhat-0}
 \hat{v}(0) = \int_{\R^\nu}\! \mathrm{d}x\ v(x) >0\quad\text{and,}
 \quad \hat{v}(0) \geq \hat{v}(q), \quad \forall q \in \R^\nu,
\end{equation}
where $\hat{v}(q) \geq 0$ by (\textit{a}). 
\end{enumerate}
Notice that by virtue of (\textit{a})
the potential $v(x)$ at $x=0$ is finite and that $v(0) \geq v(x)$, 
$\forall x \in \R^\nu$.

It is shown \cite{ruelle:1969} that under conditions \ab\ the 
interaction is superstable, \ie the $n$-body potential satisfies 
the inequality
\begin{equation*}
\sum_{1 \leq i < j \leq n}v(x_i - x_j) \geq 
\frac{A}{2V}n^2 - B n
\end{equation*}
for some constants $A > 0$, $B \geq 0$,  for all $n \in \N$, $x_i \in \L$ and 
$\L$ large enough which implies that the
thermodynamic potentials exist for all values of the chemical potential 
$\mu$. Moreover, our proof requires a more stringent condition: 
it asks the constant $A = \hat{v}(0)(1-\epsilon)$, with $\epsilon > 0$
arbitrarily small and $B = v(0)/2$. These \textit{optimal stability constants} 
for the $L^1$-integrable potentials of positive type were established by Lewis, 
Pul\`e and de Smedt \cite{lewis:1984}. Since these optimal constants are 
important for our proof, we incorporate in this note the 
corresponding argument 
(see section~\ref{psp}).

The superstability of the particle interaction is, together with the assumption of a 
spectral gap (\ref{T-Delta}), the physical foundation of our proof.  
We prove that the zero-mode Bose-Einstein condensation  in the groundstate of 
the one-particle  
spectrum isolated from the continuous spectrum by a gap $\D$ is robust enough 
under the application of a superstable interaction, provided that this spectral
gap is large enough. 
Therefore, we show that in this case the zero-mode BEC is stable with 
respect to a class of superstable interactions.
More technically our main result, announced in
\cite{lauwers:2003}, can be stated as follows:

\begin{theorem}\label{theorem}
Consider a system of interacting Bose particles (\ref{H-int}) in three or more
dimensions, with a two-body interaction satisfying conditions \ab.
Fix an inverse temperature and a chemical potential $(\beta,\mu)$
such that $\mu > g\hat{v}(0)\rho^{P}_c(\beta)$, then there exists 
a minimal value for the gap $\D_{min}$, such that for all $\Delta \geq \D_{min}$
the thermodynamic limit ($\lim_\L :$ $V\to\infty$) of the $k=0$ mode particle
number occupation density is positive:
\begin{equation*}
\rho_{0,g}^{\D}(\beta,\mu) \equiv
\lim_{\L} \frac{1}{V}\langle N_0 \rangle_{H_{\L,g}^\D}(\beta,\mu)  > 0.
\end{equation*}
$\langle - \rangle_{H_{\L,g}^\D}(\beta,\mu)$ denotes the grand-canonical Gibbs 
state for the model $H_{\L,g}^\D$ \eqref{H-int} at inverse temperature $\beta =
1/k_BT$ and chemical potential $\mu$. $\hat{v}(0)$ is as in \eqref{vhat-0} 
and $\rho^{P}_c(\beta)$ is the critical density of the perfect Bose gas at 
inverse temperature $\beta$.
\end{theorem}
 A detailed proof is found in 
section~\ref{proofs}, Figures~\ref{fig2}--\ref{fig4} illustrating this theorem can be 
found in section~\ref{disc}. 
Notice that an analogous result as in Theorem~\ref{theorem} holds in 
lower dimensions $\nu \leq 2$ (cf.~section~\ref{disc}). Our proof of BEC is 
based on comparing the condensate density of the full model \eqref{H-int} with 
the condensate density of specially tuned reference systems.
This yields (various) lower bounds on the condensate of the full system in 
terms of thermodynamic potentials of the reference systems. 
The reference systems we use are based on the mean-field Bose gas. Its 
thermodynamic properties, necessary in our proof of BEC, are reviewed in 
section~\ref{mf}. 

This paper is organised as follows: in section~\ref{prel} the essential 
preliminaries, \ie superstable potentials and the thermodynamics of the 
mean-field Bose gas, are reviewed, section~\ref{proofs} contains the actual 
proofs of the new results and a discussion on these results 
(section~\ref{disc}) concludes the paper.

\section{Preliminaries \label{prel}}

\subsection{Superstable Potentials} \label{psp}

An important aspect of particle interactions in both classical and quantum 
continuous systems are their stability properties
\cite{ruelle:1969,bratteli:1996,zagrebnov:2001}. 
If the particle interaction is such that the grand-canonical partition
function does not converge, it is called catastrophic. 
It means that the thermodynamic pressure is not everywhere well defined, 
and good thermodynamic behaviour is excluded in those domains. 

A useful criterion for stability is the following: a pair-potential $v$ is 
called stable \cite{ruelle:1969} if the corresponding $n$-particle 
interactions can be estimated from below as :
\begin{equation}
\sum_{1 \leq i < j \leq n}v(x_i - x_j) \geq - B n,
\end{equation}
for a $B \geq 0$, and all $n \geq 2$, $x_i \in \L$.
This is a sufficient condition for good thermodynamic behaviour. 
An important subclass of the stable interactions are the so-called 
superstable pair-potentials \cite{ruelle:1969}, satisfying 
\begin{equation}\label{s-stable}
\sum_{1 \leq i < j \leq n}v(x_i - x_j) \geq 
\frac{A}{2V}n^2 - B n
\end{equation}
for some constants $A > 0$, $B \geq 0$, and all $n \geq 2$, $x_i \in \L$, 
where $\L$ is large enough. This condition (\ref{s-stable}) yields the 
existence of the grand-canonical pressure for all values of the chemical 
potential $\mu \in \R$.
From \eqref{s-stable}, it follows that the interaction term 
(\ref{U-Lambda}) satisfies the operator inequalty :
\begin{equation}\label{sstab2}
U_\L \geq \frac{A}{2V}N_\L^2 - B N_\Lambda. 
\end{equation}
An example of superstable potentials are those verifying the conditions \ab.
For this class of potentials Lewis, Pul\`e, and de Smedt \cite{lewis:1984} proved 
the existence of optimal constants $A = \hat{v}(0)(1 - \epsilon)$ and 
$B = v(0)/2$ in (\ref{s-stable}). Here, $\epsilon > 0$ is a  positive constant
related to the volume of $\L$. Since we use this result, we give a 
version of their proof adapted to our situation.

\begin{lemma}[Lewis, Pul\`e, and de Smedt \cite{lewis:1984}]\label{lps-lemma}
Take $\epsilon > 0$, and take a  real $L^1$-function of positive type $v
: \R^\nu \to \R$ with $\hat{v}(0) > 0$ (\ref{vhat-0}).
There exist a subset $\L_{min} \subset \R^\nu$ such that 
for each open box $\L$, with $\L_{min} \subset \L$, the 
following inequality holds
\begin{equation}\label{lps}
\sum_{1 \leq i < j \leq n}v(x_i - x_j) \geq 
-\frac{v(0)}{2}n + \frac{\hat{v}(0)}{2V}(1 -\epsilon)n^2, 
\end{equation} 
for all $n \geq 2$, and each set of $n$ distinct points 
$\{x_1,\ldots,x_n\} \subset \L$.
\end{lemma}

\textit{Proof.} By Bochner's theorem \cite{reed:1975} a function of positive
type $v$ defines a positive-definite quadratic form 
$\langle \cdot,\cdot\rangle_v$ on the 
space of bounded measures on $\R^\nu$ :
\begin{equation*}
\langle \mu_1, \mu_2 \rangle_v = \iint_{\R^\nu \times \R^\nu} \!
v(x -y)\ \mathrm{d}\mu_1(x)
\mathrm{d}\mu_2(y).
\end{equation*} 
Hence, by the Cauchy-Schwarz inequality this quadratic form satisfies 
\begin{equation*}
\langle \mu_1, \mu_1 \rangle_v \geq \frac{|\langle \mu_2, \mu_1 \rangle_v|^2}
{\langle \mu_2, \mu_2 \rangle_v}.
\end{equation*}
 Applying this inequality to the measures
\begin{equation*}
\mathrm{d}\mu_1(x) = \sum_{i=1}^n\delta_{x_i}(x)\mathrm{d}x
\ \text{and}\ \mathrm{d}\mu_2(x) = \chi_{B(\L,h)}(x)\mathrm{d}x, 
\end{equation*}
where $\chi_{B(\L,h)}(x)$ is the indicator of the set 
$B(\L,h) = \{ x \in \R^\nu : \text{dist}(x,\L) \leq h \}$, and 
$x_i \in \L, i = 1,\ldots,n$, $n > 1$, we arrive at the following estimate:
\begin{equation}\label{lps1}
\sum_{i,j =1}^n v(x_i -x_j) \geq \frac{\left| \sum_{i =1}^n A_\L^h(x_i)\right|^2}
{\int_{B(\L,h)}\mathrm{d}y\ A_\L^h(y)},
\end{equation}
where
\begin{equation*}
A_\L^h(y) = \int_{B(\L,h)}\!\mathrm{d}x\ v(x -y).
\end{equation*}
A lower bound for the numerator in the r.h.s. of \eqref{lps1} is found using 
that for all $y \in \L$ one has : 
\begin{align}\nonumber
\left |A_\L^h(y) - \hat{v}(0)\right|  & \leq  \left | A_\L^h(y)
- \int_{|x-y|\leq h}\mathrm{d}x\ v(x-y)\right| + \int_{|x|> h}\mathrm{d}x\ |v(x)|
\\ & \leq \delta(h), \label{teller}
\end{align}
where $\delta(h)$ is given by:
\begin{equation}\label{deltah}
\delta(h) = 2 \int_{|x|> h}\mathrm{d}x\ |v(x)|.
\end{equation}
The last step in \eqref{teller} is valid since for every $y \in \L$, there is a
ball with the radius $h$ and the centre at $y$ lying inside $B(\L,h)$.
Since $v$ is an $L^1$-function, $\delta(h)$ \eqref{deltah} converges to zero in
the limit $h \to \infty$. This yields
\begin{equation*}
A_\L^h(y) \geq \hat{v}(0) - \delta(h) > 0, 
\end{equation*}
for all $ y \in \L$ and for $h$ large enough. The last bound is valid 
since $\hat{v}(0) > 0$, and since $\delta(h)\geq 0$ \eqref{deltah} 
becomes arbitrarily small for $h$ large enough. 
The numerator in the r.h.s.\ of \eqref{lps1} is therefore bounded from below as 
follows : 
\begin{equation}\label{lps2}
\left| \sum_{i =1}^n A_\L^h(x_i)\right|^2 \geq 
n^2 \left( \hat{v}(0) - \delta(h)\right)^2.
\end{equation}
An upper bound for the denominator in the r.h.s. of \eqref{lps1} is based on the
following estimate
\begin{align}\nonumber
\left|\int_{B(\L,h)}\! \mathrm{d}y\ A_\L^h(y) - V \hat{v}(0) \right| & 
\leq \left|\int_{B(\L,h)}\!\mathrm{d}y\ A_\L^h(y)- \int_{\L}\!\mathrm{d}y\ 
A_\L^h(y)\right| + \left| \int_{\L}\!\mathrm{d}y\ A_\L^h(y)
- V \hat{v}(0) \right| \\ \label{lps3}
& \leq \text{vol}(B(\L,h)\backslash \L)\|v\|_{L^1} + V\delta(h),
\end{align}
see \eqref{teller}.
Hence, using the estimates \eqref{lps2} and \eqref{lps3} in \eqref{lps1}, 
one gets
\begin{equation*}
\sum_{i,j =1}^n v(x_i -x_j) \geq \frac{n^2}{V}
\frac{\left( \hat{v}(0) - \delta(h)\right)^2}
{\hat{v}(0) + \|v\|_{L^1}\text{vol}(B(\L,h)\backslash \L)/V + \delta(h)}.
\end{equation*}
Notice that for any smooth-shaped $\L$ the factor 
$\text{vol}(B(\L,h)\backslash \L)/V$ is of the order $\order{h/L}$,
and vanishes in the limit $V \to \infty$ for fixed $h$. Since 
\begin{equation*}
\frac{\left( \hat{v}(0) - \delta(h)\right)^2}
{\hat{v}(0) + \|v\|_{L^1} \text{vol}( B(\L,h)\backslash \L)/V + \delta(h)}
\geq \hat{v}(0) - \|v\|_{L^1} \text{vol}( B(\L,h)\backslash \L)/V - 3\delta(h),
\end{equation*}
we can choose $h$ large enough, such that
$\delta(h)/\hat{v}(0) < \epsilon/4$, and then take $\L_{min}$ 
that $\|v\|_{L^1} \text{vol}( B(\L,h)\backslash \L)/ V \hat{v}(0) < \epsilon/4$,
to obtain the estimate :
\begin{equation*}
\sum_{i,j =1}^n v(x_i -x_j) \geq \frac{n^2}{V}\hat{v}(0)(1 -\epsilon),
\end{equation*} 
for all boxes $\L$ with $\L_{min} \subset \L$. \hfill $\square$

As it follows from the proof, this result holds for more general shapes 
of $\L$, the only condition is that $\text{vol}(B(\L,h)\backslash \L)/V$ 
tends to zero in the limit $V \to \infty$ for some fixed $h>0$. 
In fact, $\L$ tends to $\R^\nu$ in the sense of Van Hove \cite{ruelle:1969} 
is enough.

\begin{remark} \label{remark}
From the proof it also follows that  the value of
$\epsilon $ is defined
by $\Lambda_{min}$, and that increasing the latter we can make  $\epsilon $
as small as we want. This means that \emph{after} the thermodynamic limit
one can put $\epsilon = 0$.
\end{remark}

In \cite{lewis:1984}, it was also shown that the constants 
$A =\hat{v}(0)(1 -\epsilon)$ and $B = v(0)/2$ in Lemma~\ref{lps-lemma} are 
optimal for this type of pair-potentials. This is based on the following
argument: suppose there exists a series of positive constants $\{A_l\}_l$ 
converging for $l \to \infty$ to a better superstability constant 
$A > \hat{v}(0)$, \ie such that for all $l$
\begin{equation}\label{opt1}
\sum_{1\leq i < j \leq n }v(x_i - x_j) \geq \frac{n^2}{V_l}A_l,
\end{equation}
for all finite sets of distinct points $\{x_1,\ldots ,x_n\}\subset \L_l$. 
Since we can choose  $\epsilon>0$ small enough that $A - \epsilon >
\hat{v}(0)$, there exists $l_{1}(\epsilon)$ such that $A_l > 
A - \epsilon /2$ for all $l>l_{1}(\epsilon)$.
On the other hand, for all $\L_l$ large enough, $l>l_2(\epsilon)$, we get 
\begin{equation*}
\int_{\L_l} \! \mathrm{d}x \ v(x-y) \leq  \hat{v}(0) + \epsilon/2,
\end{equation*}
uniformly in $y \in \R^\nu$.
Then by integration of both sides of \eqref{opt1} over $\L_l^n$, for 
$l> max(l_{1}(\epsilon),l_{2}(\epsilon))$ we get
the estimates : 
\begin{equation*}
nv(0) + \frac{n(n-1)}{V_l} (\hat{v}(0) + \epsilon/2)  
\geq 
\frac {1}{V_{l}^{n}}\int_{{\L_l}^n} \! \mathrm{d}x_1...\mathrm{d}x_n\ 
\sum_{i,j}v(x_i - x_j) 
\geq \frac{n^2}{V_l} (A - \epsilon /2).
\end{equation*}
Since above the $n>1$ is arbitrary, these estimates imply 
that $A - \epsilon \leq \hat{v}(0)$,
which is in contradiction to the hypothesis, hence yielding the optimality
of the constants in Lemma~\ref{lps-lemma}.

\subsection{Thermodynamics of the Mean-Field Bose Gas} \label{mf}

In this section we briefly review the properties of the so-called mean-field 
Bose gas (sometimes also called the imperfect Bose gas), an exactly solvable 
model of Bosons \cite{huang:1967, davies:1972, fannes:1980b, buffet:1983,
berg:1984,papoyan:1986, lewis:1988} (for an extended review see 
\cite{zagrebnov:2001}), which will play the r\^ole of a reference system. 
The mean-field Bose gas is defined by the local Hamiltonians
\begin{equation}\label{H-MF}
H^{\Delta}_{\Lambda,\l} = T_\Lambda^\Delta +
\frac{\l}{2V}N_\Lambda^2.
\end{equation}
The kinetic energy operator $T_\Lambda^\Delta$  (\ref{T-Delta}) is identical to
the one of the  fully interacting system (\ref{H-int}), but the interaction 
term (\ref{U-Lambda}) is replaced by a kind of mean-field interaction term.
The physical relation of the model \eqref{H-MF} to our system \eqref{H-int} 
lies in the fact that it is the Van der Waals limit of the fully interacting 
system \eqref{H-int} \cite{buffet:1983b}. In that case, the constant $\l$ has to be 
chosen equal to the long-range part of the interaction \eqref{U-Lambda}, \ie 
$\l = g\hat{v}(0)$.  

The explicit solution for the gapless case $\D =0$ of this model can be
found at several places \cite{huang:1967, davies:1972, fannes:1980b, 
buffet:1983,berg:1984,papoyan:1986, lewis:1988,zagrebnov:2001}. Here, we focus 
on the case with non-vanishing gap. The grand-canonical pressure function at 
inverse temperature $\beta$ and chemical potential $\mu$ is defined as
\begin{equation*}
p_{\L,\lambda}^\D(\beta,\mu) = \frac{1}{\beta V}\ln\Tr_{\mathcal{F}_B} 
\e^{-\beta(H^{\Delta}_{\L,\lambda} -\mu N_\L )}.
\end{equation*}
$\Tr_{\mathcal{F}_B}$ denotes the trace over the boson Fock space 
$\mathcal{F}_B$. Below, we develop explicit expressions for the pressure and the 
particle densities in the thermodynamic limit $V \to \infty$.  

\begin{lemma}[Thermodynamic Functions]
The grand-canonical pressure $p_{\l}^\D(\beta,\mu) = 
\lim_{\L }p_{\L,\lambda}^\D(\beta,\mu)$
of the mean-field Bose gas \eqref{H-MF}
exists for all $\beta \geq 0, \mu \in \R$ and is given by
the Legendre transform:
\begin{equation}\label{mfp}
p_{\l}^\D(\beta,\mu) = \sup_{\rho \geq 0}\left(\mu\rho - 
f_\l^\D(\beta,\rho)\right),
\end{equation}
where the canonical free energy $f_\l^\D(\beta,\rho)$ at inverse temperature 
$\beta$ and density $\rho$ is given by
\begin{equation}\label{mff}
f_\l^\D(\beta,\rho) = f^{P,\D}(\beta,\rho) + \frac{\l}{2} \rho^2, 
\end{equation}
$f^{P,\D}(\beta,\rho)$ is the free energy of the perfect Bose gas 
with gap $\D$ \eqref{T-Delta}.
\end{lemma}
\textit{Proof.}
The thermodynamic pressure of the perfect Bose gas is given by:
\begin{equation*}
p^{P,\D}(\beta,\mu) = \lim_{\L } p^{P,\D}_\L(\beta,\mu) =
\lim_{\L }\frac{1}{\beta V}\Tr_{\mathcal{F}_B} 
\e^{-\beta(T^{\Delta}_{\L} -\mu N_\L )}
\end{equation*}
which implies that in order to be well defined, $\mu$ must be bounded from
above: $\mu < -\D$, \ie
\begin{equation*}
p^{P,\D}_\L(\beta,\mu) = \frac{1}{\beta V}\ln \sum_{n_{0}=0}^{\infty}
\e^{\beta(\D + \mu)n_0} \sum_{\{n_k\}_{k \ne 0}}\e^{-\beta(\varepsilon_k 
-\mu)n_k}.
\end{equation*}
The canonical free energy $f^{P,\D}(\beta,\rho)$, is the Legendre 
transform of $p^{P,\D}(\beta,\mu)$, defined only for $\mu \leq -\D$,
\begin{equation}\label{Pf}
f^{P,\D}(\beta,\rho) =
\sup_{\mu \leq -\D}\left(\rho\mu - p^{P,\D}(\beta,\mu)\right). 
\end{equation}
By direct calculation one finds expression \eqref{mff} for
the free energy of the mean-field model \eqref{H-MF}
at temperature $\beta$ and density $\rho$ as
\begin{equation*}
f_\l^\D(\beta,\rho) = \lim_{\L } -\frac{1}{\beta V}\ln\Tr_{\mathcal{H}_B^{(n)}} 
\e^{-\beta H^{\Delta}_{\L,\lambda}},
\end{equation*}
where $\Tr_{\mathcal{H}_B^{(n)}}$ denotes the trace over the Hilbert space 
$\mathcal{H}_B^{(n)}$ of
symmetrised functions for $n =  \lfloor \rho V \rfloor$ (integer part of 
$\rho V$) Bosons. 
Since on this space the mean-field interaction term is constant, we 
immediately find :
\begin{equation*}
\lim_{\L }f_\L[H_{\L,\l}^\D](\beta,\rho) = \lim_{\L } f_\L[T^\D_\L](\beta,\rho) +
\frac{\l}{2}\rho^2.
\end{equation*}
The pressure of the mean-field gas, is again the Legendre 
transform of $f_\l^\D(\beta,\rho)$, yielding formula \eqref{mfp}, 
well defined for all $\mu \in \R$. 
\hfill $\square$

\begin{theorem}[Pressure of the Mean-Field Bose Gas]\label{t-p-MF}
The grand-canonical pressure of the mean-field Bose Gas \eqref{H-MF}
is explicitly given by
\begin{equation}\label{p-MF}
p^\D_\l(\beta,\mu) = \left\{ 
\begin{array}{lr}
p^{(\D=0)}_\l(\beta,\mu), 
&\text{for}\quad 
\mu \leq - \D + \l \rho^P(\beta,-\D);
\\[0.2cm]
 (\mu + \D)^2/2\l + p^P(\beta,-\D),
 & \text{for}\quad
\mu  > -\D + \l \rho^P(\beta,-\D),
\end{array} \right.
\end{equation}
where $p^P(\beta,\mu)$, and $\rho^P(\beta,\mu)$ are respectively the pressure,
and the total density of the perfect Bose gas; $p^{(\D=0)}_\l(\beta,\mu)$ 
is the pressure of the mean-field Bose gas without gap.
\end{theorem}
\textit{Proof.}
The formula \eqref{p-MF} is found by working out explicitly the Legendre
transforms in \eqref{mfp}--\eqref{Pf}, and using the properties of the
perfect Bose gas.
Let $\mu = \bar{\mu}_\L(\beta,\rho)$ be solution of the equation:
\begin{equation*}
\rho = \frac{1}{V}\langle N_\L \rangle_{T_\L^{\D}}
(\beta,\bar{\mu}_\L(\beta,\rho)),
\end{equation*}
for a given $\rho$ and $\beta$, where the right-hand side is the 
expectation value of the total density in the grand-canonical Gibbs
state for the perfect Bose gas model $T_\L^{\D}$.
Denote the limiting solution by 
\begin{equation*}
\bar{\mu}(\beta,\rho) = \lim_{\L} \bar{\mu}_\L(\beta,\rho),
\end{equation*}
We have $\bar{\mu}(\beta,\rho) \leq -\D$, if $\rho \leq  
\rho^P(\beta,-\D)$, and $\bar{\mu}(\beta,\rho) = -\D$, if $\rho \geq 
\rho^P(\beta,-\D)$, hence
\begin{equation}\label{Pf2}
f^{P,\D}(\beta,\rho) = \left\{ 
\begin{array}{ll}
\rho\bar{\mu}(\beta,\rho) - 
p^P(\beta,\bar{\mu}(\beta,\rho)),&\text{if}\quad \rho \leq  \rho^P(\beta,-\D);\\[0.2cm]
-\rho\D - p^P(\beta,-\D),&\text{if}\quad \rho >  \rho^P(\beta,-\D).
\end{array} \right.
\end{equation}
This is the explicit expression for \eqref{Pf}. 
The thermodynamic potentials such as the pressure and the particle density of 
the free Bose gas with gap $\D$ are the same as for the gapless perfect Bose 
gas, but only for the 
values of $\mu < -\D$. At $\mu = -\D$, there is degeneracy of the densities 
and BEC occurs.
Since for $\D > 0$ the critical density 
$\rho_c^{P,\D}(\beta) \equiv \rho^P(\beta,-\D)$ is 
finite in all dimensions $\nu \geq 1$, the condensation takes place in all 
dimensions whereas in the gapless case
$\rho_c^{P,(\D=0)}(\beta) \equiv \rho_c^{P}(\beta) < \infty$ 
only for dimensions 
$\nu > 2$. Hence condensation occurs only in three or more dimensions 
at $\beta < \infty$. 

Recall now the expression for the canonical free energy of the mean-field Bose
gas, Eq.~\eqref{mff}. Using
the expression for the free energy of the perfect Bose gas $f^{P,\D}(\beta,\rho)$
\eqref{Pf2} derived above, one finds for $\partial_\rho f^\D_\l(\beta,\rho)$,
\begin{equation}\label{drhomff}
\partial_\rho f^{\D}_\l(\beta,\rho) = \left\{ 
\begin{array}{ll}
\partial_\rho f^{P}(\beta,\rho) + \l\rho, & \text{if}\quad
\rho \leq  \rho^P(\beta,-\D); 
\\[0.2cm]
-\D + \l\rho,& \text{if}\quad \rho >  \rho^P(\beta,-\D).
\end{array} \right.
\end{equation}
By virtue of \eqref{drhomff} and \eqref{mfp}, one gets the expression for the 
the mean-field Bose gas pressure: 
\begin{multline*}
p^\D_\l(\beta,\mu) = \sup_{\rho \geq 0}\left(\mu\rho - 
f_\l^\D(\beta,\rho)\right) \\[0.2cm]
= \left\{ 
\begin{array}{ll}
\mu \bar{\rho}(\beta,\mu) - f_\l^\D(\beta,\bar{\rho}(\beta,\mu)), 
&\text{for}\quad 
\mu \leq - \D + \l \rho^P(\beta,-\D);
\\[0.2cm]
 \mu (\mu + \D)/\l - f_\l^\D(\beta,(\mu+\D)/\l),
 & \text{for}\quad
\mu  > -\D + \l \rho^P(\beta,-\D),
\end{array} \right.
\end{multline*}
where $\bar{\rho}(\beta,\mu)$ is the solution of
$\partial_\rho f^{P}(\beta,\bar{\rho}(\beta,\mu)) + 
\l \bar{\rho}(\beta,\mu) = \mu $ as a function of 
$\mu \leq -\D + \l\rho^P(\beta,-\D)$ and $\beta$.
Since by \eqref{Pf2}
\begin{equation*}
f^{P,\D}(\beta,(\mu+\D)/\l) = -\D \frac{\mu +\D}{\l}-
p^P(\beta,-\D),
\end{equation*}
for $\mu+\D > \l \rho^P(\beta,-\D)$, where $p^P(\beta,-\D)$ is the pressure of
the free Bose gas, we use the expression \eqref{mff} for the free 
energy of the mean-field gas to find \eqref{p-MF}, that proves 
Theorem~\ref{t-p-MF}. \hfill $\square$

\begin{theorem}\label{MF-densities}
Considering the mean-field Bose gas \eqref{H-MF}, we derive the following
expressions for the densities in the thermodynamic limit.
The total grand-canonical density is given by
\begin{equation}\label{rho-MF}
\rho^\D_\l(\beta,\mu) = \left\{ 
\begin{array}{ll}
\rho^{(\D=0)}_\l(\beta,\mu), 
&\text{for}\quad 
\mu \leq - \D + \l \rho^P(\beta,-\D);
\\[0.2cm]
 (\mu + \D)/\l,
 & \text{for}\quad
\mu  > -\D + \l \rho^P(\beta,-\D).
\end{array} \right.
\end{equation}
The zero-mode condensate density is given by
\begin{equation}\label{rho0-MF}
\rho^\D_{0,\l}(\beta,\mu) = \left\{ 
\begin{array}{ll}
0, 
&\text{for}\quad 
\mu \leq - \D + \l \rho^P(\beta,-\D);
\\[0.2cm]
 (\mu + \D)/\l - \rho^P(\beta,-\D),
 & \text{for}\quad
\mu  > -\D + \l \rho^P(\beta,-\D).
\end{array} \right.
\end{equation}
The limit of the expectation value $\langle N_\L^2
\rangle_{H^\D_{\L,\l}}(\beta,\mu)/V^2$ is 
given by
\begin{equation}\label{rho2-MF}
\lim_{\L}\frac{1}{V^2}\langle N_\L^2
\rangle_{H^\D_{\L,\l}}(\beta,\mu) = \left\{ 
\begin{array}{ll}
\rho^{(\D=0)}_\l(\beta,\mu)^2, 
&\text{for}\quad 
\mu \leq - \D + \l \rho^P(\beta,-\D);
\\[0.2cm]
 (\mu + \D)^2/\l^2,
 & \text{for}\quad
\mu  > -\D + \l \rho^P(\beta,-\D).
\end{array} \right.
\end{equation}
Here $\rho^{(\D=0)}_\l(\beta,\mu)$ is the density for the gapless 
mean-field gas, \ie for $\D=0$ in Eq.~\eqref{H-MF}. 
\end{theorem}
\textit{Proof.}
These quantities are derived using that the pressure \eqref{p-MF} is a convex
function of respectively \ $\D,\mu$ and $\l$. By the Griffith lemma 
\cite[Appendix C]{zagrebnov:2001}, the order of the 
thermodynamic limit and the corresponding derivative 
can be interchanged, which gives :
\begin{gather*}
\rho_{0,\l}^{\D}(\beta,\mu) = \lim_{\L}\frac{1}{V}\langle N_0 
\rangle_{H^\D_{\L,\l}}(\beta,\mu) =
\lim_{\L} \partial_\D p^\D_{\L,\l}(\beta,\mu) = \partial_\D
p^\D_{\l}(\beta,\mu); \\
\rho_{\l}^{\D}(\beta,\mu) = \lim_{\L}\frac{1}{V}\langle N_\L 
\rangle_{H^\D_{\L,\l}}(\beta,\mu) =
\lim_{\L} \partial_\mu p^\D_{\L,\l}(\beta,\mu) = \partial_\mu
p^\D_{\l}(\beta,\mu); \\
\lim_{\L} \frac{1}{V^2}\langle N_\L^2 
\rangle_{H^\D_{\L,\l}}(\beta,\mu) = 
\lim_{\L} - 2 \partial_\l p^\D_{\L,\l}(\beta,\mu) =
 - 2 \partial_\l p^\D_{\l}(\beta,\mu).
\end{gather*}
By virtue of \eqref{p-MF} of Theorem~\ref{t-p-MF} these imply the 
explicit expressions 
\eqref{rho-MF}--\eqref{rho2-MF} of Theorem~\ref{MF-densities}. \hfill $\square$

Taking the limit $\D \downarrow 0$, we recover the usual expressions for the
mean-field Bose gas \eqref{H-MF} with vanishing gap, in particular the
expression for the zero-mode condensate density in dimensions $\nu > 2$,
\begin{equation}\label{rho0-MF-0}
\rho^{(\D=0)}_{0,\l}(\beta,\mu) = \left\{ 
\begin{array}{ll}
0, 
&\text{for}\ 
\mu \leq  \l \rho^P_c(\beta);
\\[0.2cm]
 \mu/\l - \rho^P_c(\beta),
 & \text{for}\
\mu  > \l \rho^P_c(\beta).
\end{array} \right.
\end{equation}

\section{Proofs of the Main Results \label{proofs}}
The main idea of the proof of the condensation for the systems \eqref{H-int} 
is to estimate their Bose condensate from below by the condensate of a 
particularly chosen reference system for which one can compute the 
amount of the condensate explicitly. Thus, a judicious choice of this reference 
system is a subtle point of our proof. 

Since we consider superstable systems, \ie systems where the grand-canonical 
pressure is  defined for all values of the chemical potential, it seems to be 
natural to choose a reference system which is also superstable. This 
immediately rules out the perfect Bose
gas \eqref{T-Delta} as a reference system, since its pressure is only 
well defined for $\mu \leq -\D$. Choosing the reference systems within the 
class of
mean-field Bose gases (cf.~section~\ref{mf}), which are indeed well-known 
superstable systems, seems therefore a good choice. 
The reference systems that we consider are mean-field Bose systems which
are \textit{close enough} to the Van der Waals limit of the fully interacting
system \eqref{H-int}. 
Apart from the use of a reference system, the proof is based 
on various convexity properties of the thermodynamic functions. In particular 
it is based on the following lemma.

\begin{lemma}\label{lemma-lb}
The zero-mode condensate density $\rho^\Delta_{0,g}(\beta,\mu)$ in the 
thermodynamic limit
of grand-canonical Gibbs states of interacting system (\ref{H-int}) with a
superstable two-body potentials $v$ satisfying the conditions \ab, has the 
following lower bound :
\begin{equation}\label{lb}
\begin{split}
\rho^\Delta_{0,g}(\beta,\mu) \geq &\ \frac{\mu}{g\hat{v}(0)} +
\frac{g\hat{v}(0)}{2\Delta}\rho^{P}(\beta,-\Delta)^2 
- \frac{g v(0)}{2\Delta}\rho^{(\Delta = 0)}_g(\beta,\mu)
\\
& -  \frac{\mu +\Delta}{\Delta}\rho^{P}(\beta,-\Delta)  
 - \rho^{P}_c(\beta).
\end{split}
\end{equation}
Here $\rho_g^{(\Delta = 0)}(\beta,\mu)$ denotes the total
density of the interacting gas without gap \eqref{H-int}. 
$\rho^{P}(\beta,-\Delta)$ refers to the total 
density of the perfect Bose gas at the inverse temperature $\beta$ and the 
chemical 
potential $\mu = -\D$, $\rho_c^{P}(\beta)$ is the critical density of the 
perfect Bose gas. The bound is valid for values $\mu > g\hat{v}(0)
\rho_c^{P}(\beta)$, and dimensions $\nu > 2$.
\end{lemma}

\textit{Proof}.
The pressure $p_\L[H^\D_\L]$ 
of systems with a gap in the kinetic energy spectrum \eqref{T-Delta}
and any stable interaction is an increasing convex function of the 
parameter $\D \geq 0$. 
Since by Theorem~\ref{MF-densities} the condensate density 
$\rho^\Delta_{0,g}(\beta,\mu)$ is the derivative of the corresponding 
pressure with respect to $\D$, the 
convexity property yields a lower bound for the zero-mode density 
$\langle N_0/V \rangle_{H_{\L, g}^{\D}}$ : 
\begin{equation}\label{c1}
\frac{1}{V}\langle N_0 \rangle_{H_{\L, g}^{\D}} 
\geq  \frac{p_\L[H^\D_{\L,g}] - 
p_\L[H^{(\D =0)}_{\L,g}]}{\D}.
\end{equation}

Now we use a reference system to get the lower bound on the condensate. 
This reference system is a mean-field Bose gas \eqref{H-MF}, 
defined by the local Hamiltonian (cf.~section~\ref{mf}) 
\begin{equation}\label{H-MF2}
H^{\D}_{\L,g,A} = T_\Lambda^\Delta  -\mu N_\L +
g\frac{A}{2V}N_\Lambda^2.
\end{equation}
Therefore, we fix the mean-field
Bose gas \eqref{H-MF} interaction parameter by taking $\l = gA$, 
where $g > 0$ is the coupling constant
(cf.~\eqref{H-int}) and $A = \hat{v}(0)(1-\epsilon)$ is the optimal
superstability constant \eqref{s-stable} associated with the two-body 
interaction \eqref{U-Lambda} of the full model \eqref{H-int}.
By virtue of the same convexity property as in \eqref{c1}, 
the difference of the pressures between the reference Bose system \eqref{H-MF2} 
with gap and without gap, is bounded from below by the condensate density 
for the reference mean-field gas without gap, \ie
\begin{equation}\label{c2}
\frac{p_\Lambda[H^{\D}_{\L,g,A}] - p_\L[H^{(\D =0)}_{\L,g,A}]}{\D} \geq 
\frac{1}{V}\langle N_0 \rangle_{H_{\L,g,A}^{(\D=0)}} .
\end{equation}
Adding the inequality \eqref{c2} to the lower bound on the
condensate density of the full system \eqref{c1}, we introduce the reference
system \eqref{H-MF2} in our estimate:
\begin{equation}\label{lb1}
\frac{1}{V}\langle N_0 \rangle_{H_{\L, g}^{\D}} 
\geq \frac{1}{V}\langle N_0 \rangle_{H_{\L,g,A}^{(\D=0)}} 
- \frac{1}{\D}\left(
p_\Lambda[H^{\D}_{\L,g,A}] - p_\L[H^{(\D =0)}_{\L,g,A}] 
- p_\L[H^\D_{\L,g}] + p_\L[H^{(\D =0)}_{\L,g}]
\right). 
\end{equation}
Hence, the condensate density of the interacting model \eqref{H-int} with gap
$(\D > 0)$ is bounded from below by the condensate density of the mean-field 
model
\eqref{H-MF2} without gap $(\D =0)$, and a correction term proportional to  
$1/\D$ containing the pressure differences between the full system 
and the reference system.  

These pressure differences will be estimated using the Bogoliubov convexity 
inequality \cite[Appendix D]{zagrebnov:2001}. 
Applied  to the grand-canonical pressures of the mean-field 
reference Bose gas (\ref{H-MF2}) and the full model (\ref{H-int}), it gives 
\begin{equation}\label{b-conv}
\frac{g}{V}\langle W_\L^A \rangle_{H^{\D}_{\L,g}} \leq
p_\Lambda[H^{\D}_{\L,g,A}] - p_\Lambda[H^{\Delta}_{\Lambda,g}]
\leq \frac{g}{V}\langle W_\Lambda^A \rangle_{H^{\D}_{\L,g,A}},
\end{equation}
for any $\D \geq 0$.
Here the operator $W_\L^A$ is the difference 
between the interactions of the fully interacting and the mean-field Bose 
gases : $W_\Lambda^A = U_\Lambda - A N_\L^{2}/2V$.   
Then by virtue of \eqref{lb1} and \eqref{b-conv} we get :
\begin{equation}\label{lb2}
\frac{1}{V}\langle N_0 \rangle_{H_{\L, g}^{\D}} 
\geq \frac{1}{V}\langle N_0 \rangle_{H_{\L,g,A}^{(\D=0)}} 
- \frac{g}{\D}\left(   
\frac{1}{V}\langle W_\L^A \rangle_{H^{\D}_{\L,g,A}}
-
\frac{1}{V}\langle W_\L^A \rangle_{H^{(\D = 0)}_{\L,g}} \right).
\end{equation}

Now our task is to estimate the two expectation values of $W_\L^A$ in
\eqref{lb2}.
An upper bound on $\langle W_\Lambda^A/V \rangle_{H^{\D}_{\L,g,A}}$ 
in \eqref{lb2} can be found using the properties of the pair-potential $v$
and the Gibbs states of the reference system \eqref{H-MF2}.
Expressed in terms of the creation and annihilation operators on $\L^*$, we get
 for $\langle W_\L^A/V\rangle_{H^{\D}_{\L,g,A}}$:
\begin{equation*}
\begin{split}
\frac{1}{V}\langle W_\L^A \rangle_{H^{\D}_{\L,g,A}} = 
&\ \frac{1}{2V^2} \sum_{q \in  \L^*}\sum_{k \in  \L^*} 
\sum_{k^\prime \in  \L^*} 
\hat{v}(q)
\langle a^\dagger_{k^\prime +q} a^\dagger_{k-q}
a_{k} a_{k\prime}\rangle_{H^{\D}_{\L,g,A}}
\\ & - \frac{1}{2V^2}A \sum_{k \in  \L^*} \sum_{k^\prime \in  \L^*} 
\langle a^\dagger_{k^\prime } a_{k^\prime}
a^\dagger_{k} a_{k}\rangle_{H^{\D}_{\L,g,A}}.
\end{split}
\end{equation*}
Exploiting the mode by mode gauge invariance of the Gibbs states of the
mean-field Bose gas \eqref{H-MF2}, and rewriting the above expression in terms 
of the occupation-number operators $N_k = a^\dagger_{k} a_{k}$  we arrive at
\begin{equation}\label{w-lb2}
\begin{split}
\frac{1}{V}\langle W_\L^A \rangle_{H^{\D}_{\L,g,A}} = &
\ \frac{1}{2V^2} \sum_k \sum_{k^\prime}\left(\hat{v}(0) + \hat{v}(k -k^\prime) 
- A \right)\langle N_{k} N_{k^\prime}\rangle_{H^{\D}_{\L,g,A}}
\\ & -
\frac{1}{2V^2}\hat{v}(0) \sum_k \left( \langle N^2_{k}\rangle_{H^{\D}_{\L,g,A}}
+ \langle N_{k}\rangle_{H^{\D}_{\L,g,A}}  \right).
\end{split}
\end{equation}
Since by condition (\textit{b}) :
$\hat{v}(0) \geq \hat{v}(k) \geq 0$, 
the coefficients in the first sum of the r.h.s. of \eqref{w-lb2} are 
bounded as
\begin{equation*}
\frac{1}{2}\left(\hat{v}(0) + \hat{v}(k -k^\prime) 
- A\right) \leq \hat{v}(0) - A/2.
\end{equation*}
From the second sum in the r.h.s. of \eqref{w-lb2}, we retain only 
the quadratic zero-mode term, 
by the Cauchy-Schwarz inequality we have,  
\begin{equation*}
- \frac{\hat{v}(0)}{2V^2}\langle N^2_{0}\rangle_{H^{\D}_{\L,g,A}}
\leq - \frac{\hat{v}(0)}{2V^2}\langle N_{0}\rangle^2_{H^{\D}_{\L,g,A}}.
\end{equation*}
This yields the following upper bound for  
$\langle W_\L^A/V\rangle_{H^{\D}_{\L,g,A}}$ :
\begin{equation}\label{b1}
\frac{1}{V}\langle W_\L^{A} \rangle_{H^{\D}_{\L,g,A}}
\leq  \frac{2\hat{v}(0) -A}{2V^2} \langle  N_\L^2 \rangle_{H^{\D}_{\L,g,A}}
- \frac{\hat{v}(0)}{2V^2}\langle N_0 \rangle^2_{H^{\D}_{\L,g,A}}.
\end{equation}
The expectation values appearing in the r.h.s. of
\eqref{b1} can be calculated exactly, applying  
Theorem~\ref{MF-densities}. They give in the thermodynamic limit  
the upper bound :
$(\hat{v}(0) -A/2)\rho^{\D}_{g,A}(\beta,\mu)^2 - \hat{v}(0)\rho^{\D}_{0,g,
A}(\beta,\mu)^2/2$.

The other unknown term in \eqref{lb2} is 
$\langle W_\L^A \rangle_{H^{(\D=0)}_{\L,g}}$.
It can be estimated using the superstability (\ref{sstab2}) of the 
interaction $U_\L$ (\ref{U-Lambda}) by the tuning the interaction parameter 
of the mean-field reference Bose gas (\ref{H-MF2}) to be equal to the 
constant $A$ in the superstability criterion (\ref{sstab2}), which gives
the estimate from below :
\begin{equation}\label{b2}
\frac{1}{V}\langle W_\Lambda^{A} \rangle_{H^{(\D=0)}_{\L,g}} \geq 
-\frac{B}{V}\langle N_\Lambda \rangle_{H^{(\D=0)}_{\L,g}}.
\end{equation}  
This, in particular, justifies our choice of the parameter $\l = gA$ specifying 
the reference system \eqref{H-MF2}. Using now \eqref{b1} and \eqref{b2} in 
\eqref{lb2} one finds in the thermodynamic limit 
the following lower bound for the condensate density $\rho^\D_{0,g}(\beta,\mu)$ 
\begin{equation}\label{lb3}
\begin{split} 
\rho^\D_{0,g}(\beta,\mu)  \geq & \ \rho^{(\D=0)}_{0,g,A}(\beta,\mu)
+  g\frac{\hat{v}(0)}{2\Delta}\rho^{\Delta}_{0,g,A}(\beta,\mu)^2\\
& -\frac{g}{\D}\left( B \rho^{(\Delta=0)}_{g}(\beta,\mu) +
(\hat{v}(0) - A/2)\rho^{\D}_{g,A}(\beta,\mu)^2 \right).
\end{split}
\end{equation}
The lower bound (\ref{lb}) now follows from the explicit expressions 
(Theorem~\ref{MF-densities})
for the total density and the condensate density of the mean-field 
Bose gas with gap,  
and from the well-known expression for the condensate density in the gapless
mean-field model \eqref{rho0-MF-0} for $\mu > g A\rho^{P}_c(\beta)$, \ie
in the regime where $\rho^{(\D =0)}_{0,g,A} > 0$.

In the last step to (\ref{lb}) we use the optimal superstability constants 
for continuous $L^1$-potentials of positive type 
(cf.\ Lemma~\ref{lps-lemma}), \ie we put $A = (1-\epsilon)\hat{v}(0)$, 
and $B = v(0)/2$. This gives the expression for the lower bound in the
form (\ref{lb}), since by Remark~\ref{remark}  we can put $\epsilon = 0$
after the thermodynamic limit.
\hfill $\square$

Notice that the lower bound \eqref{lb} contains the term 
$\rho^{(\Delta = 0)}_g(\beta,\mu)$, \ie the total density of the fully 
interacting gas without gap. It is not explicitly known as a function of 
$\beta$
and $\mu$. However it is always finite, and it can be viewed as a reference
parameter. 
Using a slightly modified reference system, an
alternative lower bound can be derived which consists only of explicitly known
functions related to the perfect Bose gas.
\begin{lemma}\label{lemma-lb-a}
The zero-mode condensate density $\rho^\Delta_{0,g}(\beta,\mu)$ in 
the thermodynamic limit
of the grand-canonical Gibbs states of interacting systems (\ref{H-int}) with 
superstable two-body potential $v$ satisfying conditions \ab, has the following
alternative lower bound:
\begin{equation}\label{lb-a}
\begin{split}
\rho^\Delta_{0,g}(\beta,\mu) \geq\ &  
\frac{2\mu + gv(0)}{2g\hat{v}(0)} +
\frac{g\hat{v}(0)}{2\Delta}\rho^{P}(\beta,-\Delta)^2 
   - \rho^{P}_c(\beta)
\\ &
 - \frac{2\mu + 2\D + gv(0)}{2\D \hat{v}(0)}\left(\frac{v(0)}{2}+ 
  \hat{v}(0) \rho^{P}(\beta,-\Delta)\right).
\end{split}
\end{equation}
$\rho^{P}(\beta,-\Delta)$ 
refers to the total density of the perfect Bose gas at inverse temperature 
$\beta$ and chemical potential $\mu = -\D$, and $\rho_c^{P}(\beta)$ is the 
critical density of the perfect Bose gas. The bound is valid for all values 
$\mu > g\hat{v}(0)\rho_c^{P}(\beta)$, and dimensions 
$\nu > 2$.
\end{lemma}
\textit{Proof.} The proof is completely analogous to the proof of
Lemma~\ref{lemma-lb}. But now we use the alternative reference system :
\begin{equation}\label{H-MF3}
H^{\D}_{\L,g,A,B} = T_\Lambda^\Delta  -\mu N_\L +
g\left( \frac{A}{2V}N_\Lambda^2 - B N_\L \right),
\end{equation}
which compared to the first reference system \eqref{H-MF2}, contains
an extra interaction term. Since the term  $-gBN_\L$ is 
linear in the total number operator, it corresponds to a shift in 
the chemical potential. 
Again, the constants $A$ and $B$ coincide with the optimal superstability 
values
\eqref{s-stable} for the pair-potential $v$ of the full system 
\eqref{H-int}, where $g > 0$.

First, we derive a bound similar to the one of \eqref{lb2} 
in the proof of Lemma~\ref{lemma-lb}. Now one gets : 
\begin{equation}\label{lb-a2}
\frac{1}{V}\langle N_0 \rangle_{H_{\L, g}^{\D}} 
\geq \frac{1}{V}\langle N_0 \rangle_{H_{\L,g,A,B}^{(\D=0)}} 
- \frac{g}{\D}\left(   
\frac{1}{V}\langle W_\L^{A,B} \rangle_{H^{\D}_{\L,g,A,B}}
-
\frac{1}{V}\langle W_\L^{A,B} \rangle_{H^{(\D = 0)}_{\L,g}} \right),
\end{equation}
where $W_\L^{A,B} = U_\Lambda - A N_\L^{2}/2V + B N_\L$. The expectation
values in the r.h.s. of \eqref{lb-a2} can be estimated analogously to
\eqref{b1} and  \eqref{b2}. This yields for the upper bound :
\begin{equation}\label{b-a1}
\frac{1}{V}\langle W_\L^{A,B} \rangle_{H^{\D}_{\L,g,A,B}}
\leq  \frac{2\hat{v}(0) -A}{2V^2} \langle  N_\L^2 \rangle_{H^{\D}_{\L,g,A,B}}
- \frac{\hat{v}(0)}{2V^2}\langle N_0 \rangle^2_{H^{\D}_{\L,g,A,B}}
+ \frac{B}{V} \langle  N_\L \rangle_{H^{\D}_{\L,g,A,B}}.
\end{equation}
For the lower bound of $\langle W_\L^{A,B} \rangle_{H^{(\D = 0)}_{\L,g}}$ 
we use again the superstability of the interaction $U_\L$ 
\eqref{U-Lambda}, and the fact that according to the superstability criterion 
\eqref{sstab2} we can take for $A$ and $B$ their optimal values. This gives:
\begin{equation}\label{b-a2}
\frac{1}{V}\langle W_\Lambda^{A,B} \rangle_{H^{(\D=0)}_{\L,g}} \geq 0.
\end{equation}
The explicit formula \eqref{lb-a} now follows if one introduces 
\eqref{b-a1} and \eqref{b-a2} into \eqref{lb-a2}, 
using the explicit expressions for the densities of the mean-field Bose gas
(section~\ref{mf}), and for the optimal values of $A$ and $B$ 
(section~\ref{psp}),
and finally taking  taking $\epsilon = 0$ after the thermodynamic limit, see
Remark~\ref{remark}.
\hfill $\square$

It should be remarked that one can hardly
compare the bound given in Lemma~\ref{lemma-lb} with the one in 
Lemma~\ref{lemma-lb-a},
and hence to express an opinion which of them yields the best result.
However, as the latter bound is known explicitly, it can be used to make 
numerical estimates of the condensate density and of the minimal gap as 
functions of the
various parameters involved. For this we refer to ref.~\cite{lauwers:2003} 
and section~\ref{disc}. 

Instead, we proceed now with the proof of our Theorem~\ref{theorem}, based 
on the
lower bound derived in Lemma~\ref{lemma-lb}. We prove that the condensate
density of the full model \eqref{H-int} is strictly 
positive in the domain $\mu > g\hat{v}(0)\rho^{P}_c(\beta)$ if the gap is
large enough.

\textit{Proof of Theorem~\ref{theorem}.} 
Consider the bound from Lemma~\ref{lemma-lb},
\begin{equation}\label{t1}
\begin{split}
\rho^\Delta_{0,g}(\beta,\mu) \geq &\ \ \ \frac{\mu}{g\hat{v}(0)} - 
\rho^{P}_c(\beta)
\\ & + \frac{g\hat{v}(0)}{2\Delta}\rho^{P}(\beta,-\Delta)^2
- \frac{g v(0)}{2\Delta}\rho^{(\Delta = 0)}_g(\beta,\mu)
- \frac{\mu +\Delta}{\Delta}\rho^{P}(\beta,-\Delta).
\end{split}
\end{equation}
Fix the inverse temperature $\beta$ and take the 
chemical potential $\mu$ compatible with the condition of the theorem.
This ensures the positivity of the first term in the r.h.s. of \eqref{lb3}).
Now take $\mu$ such that  
\begin{equation*}
\mu/g\hat{v}(0) - \rho^{P}_c(\beta) > 2 \eta,
\end{equation*}
for some arbitrarily chosen $\eta >0$.
This yields a lower bound for the first line in the r.h.s. of \eqref{t1}.

The expression on the second line can be absolutely bounded by $\eta$
using $\D$ large enough and the fact that 
$\lim_{\D \to \infty}\rho^{P}(\beta,-\Delta) 
= 0$. This gives :  
\begin{equation*}
\left|\frac{g\hat{v}(0)}{2\Delta}\rho^{P}(\beta,-\Delta)^2  
- \frac{g v(0)}{2\Delta}\rho_g^{(\Delta = 0)}(\beta,\mu) 
- \frac{\mu +
\Delta}{\Delta}\rho^{P}(\beta,-\Delta)\right| < \eta,
\end{equation*} 
for all $\D$ larger than some minimal gap: $\D \geq
\D_{min}$, which exists for the fixed $\mu >
g\hat{v}(0)\rho^{P}_c(\beta)$.

Collecting these two estimates, we obtain that for a fixed temperature
and $\eta > 0$ one can find $\mu$ and $\D$ large enough such that :
\begin{equation*}
\rho^\Delta_{0,g}(\beta,\mu) > \eta > 0,
\end{equation*}
proving the condensation.
\hfill $\square$

Similarly one can prove the existence of the zero-mode condensation
on the basis of the bound found in
Lemma~\ref{lemma-lb-a}.

\section{Discussion \label{disc}}
So far, we are concentrated on the case of dimensions 
$\nu > 2$,
however, the result of Theorem~\ref{theorem} can be extended to dimensions 
$\nu = 1$ or $\nu = 2$. A lower bound for the condensate density 
$\rho_{0,g}^{\D}(\beta,\mu)$ for $\nu \leq 2$ as in 
Lemma~\ref{lemma-lb} or Lemma~\ref{lemma-lb-a} is derived in a similar way. It 
requires slightly modified convexity arguments (\ref{c1})--(\ref{c2}). 
Since the free Bose gas in dimensions $\nu \leq 2$ shows only condensation in 
the case of non-vanishing gap \eqref{T-Delta}, one has to consider in 
(\ref{c1})--(\ref{c2}) the pressure differences in the form 
$p_\Lambda[H^{\D}_{\L}] - p_\L[H^{\D_0}_{\L}]$, for some $\D_0 > 0$, with 
$0 < \D_0 < \D$, instead of $\D_0 =0$. This yields the substitution in 
(\ref{lb3}) of $\rho^{(\D=0)}_{0,g,A}(\beta,\mu)$ and
$\rho_g^{(\D=0)}(\beta,\mu)$ by 
$\rho^{\D_0}_{0,g,A}(\beta,\mu)$ and
$\rho_g^{\D_0}(\beta,\mu)$.  
The bounds derived in this way are valid for all dimensions $\nu \geq 1$, and
lead to similar conclusions as in Theorem~\ref{theorem}.
Hence, in one and two dimensional interacting Bose gases with large enough
gap \eqref{T-Delta}, the zero-mode Bose-Einstein condensation is also proved.
Notice that this is in contrast 
to the Bogoliubov-Hohenberg theorem \cite{hohenberg:1967,bratteli:1996} which 
yields the absence of BEC for translation invariant continuous Bose systems 
without gap for dimensions $\nu \leq 2$.

\begin{figure}[h]
\begin{center}
\includegraphics[width=0.5\textwidth]{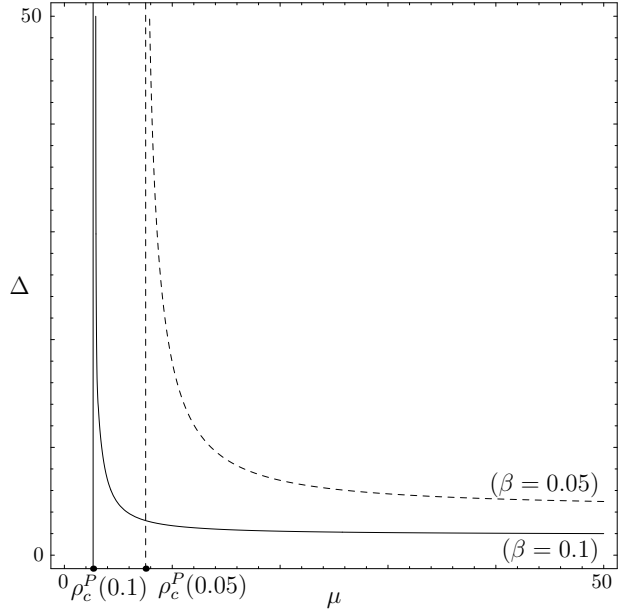}
\end{center}
\caption{\label{fig2} $(\mu,\D)$-graph with temperature dependence.}
\end{figure}

We use the lower bound \eqref{lb-a}, which can be computed explicitly as a
function of the different parameters, to visualise our estimates. 
In Figure~\ref{fig2}, the dependence of the lower bound on the temperature is
indicated.
The lines shown on the 
$(\mu,\D)$-graph indicate domains where the lower bound \eqref{lb-a} is 
positive, \ie above each of these curves we have  
BEC. The dashed curve is the threshold for condensation
calculated for inverse temperature $\beta = 0.05$, and the solid line is the 
threshold at inverse temperature $\beta = 0.1$.
Clearly, for higher values of $\beta$, the condensation occurs for smaller 
gaps and for smaller values of $\mu$, \ie at lower densities.

\begin{figure}[h]
\begin{center}
\includegraphics[width=0.5\textwidth]{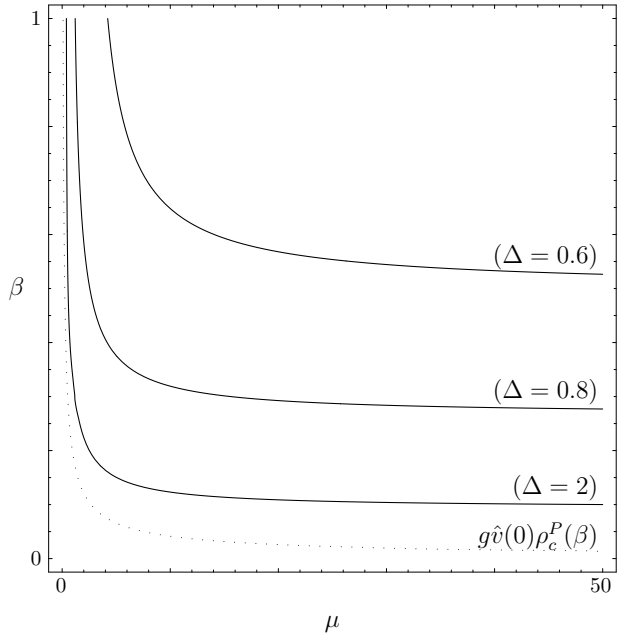}
\end{center}
\caption{\label{fig4} $(\mu,\beta)$ phase-diagram with $\D$-dependence.}
\end{figure}

To get an idea of the phase diagram of our model, on Figure~\ref{fig4} we 
present a family of thresholds as a
function of the gap value $\D$, the plotted curves are the thresholds for 
$\D = 0.6$, $0.8$, and $2$. The dotted line is the line 
$\mu = g\hat{v}(0)\rho_c^P(\beta)$, it indicates the border of validity of our
estimates.
As above, this family is calculated by equalising the lower bound
\eqref{lb-a} to zero.
Notice that to get the \emph{real} phase diagram one has to do this for the left
hand side of \eqref{lb} and not for the lower bound.

Considering the high density (large $\mu$) regime, 
the lower bound (\ref{lb-a}) can be written as
\begin{equation*}
\frac{\mu}{\hat{v}(0)} \left( \frac{1}{g} - \frac{1}{\D}\left( \frac{v(0)}{2} +
\hat{v}(0)\rho^{P}(\beta,-\Delta) \right)
\right) + o(\mu),
\end{equation*}
which means that in order to have a positive lower bound for 
$\rho^\Delta_{0,g}(\beta,\mu)$, we need
\begin{equation}\label{D-min}
\D > g\left(\frac{v(0)}{2} + \hat{v}(0)\rho^{P}(\beta,-\D)\right),
\end{equation} for high values of $\mu$. 
This means that for non-zero interaction (\ref{U-Lambda}) $(g > 0)$, there is a
non-zero lower bound on the gap width. 
Notice that the minorant \eqref{D-min} is proportional to the coupling 
constant $g \geq 0$.  

If we now choose for the two-body interaction a family of Van der Waals 
scaled pair potentials, \ie we substitute
\begin{equation}\label{subs}
v(x-y) \mapsto \l^\nu v (\l(x - y)),
\end{equation}
$\l > 0$, in the expression for the interaction term \eqref{U-Lambda}, 
then we find that 
for a fixed $\D > 0$, the condition \eqref{D-min} is satisfied if 
$\l$ is chosen small enough and for low enough temperatures.
This is easily seen as follows: substitution \eqref{subs} amounts to
substituting $\lambda^\nu v(0)$ for $v(0)$ and leaving $\hat{v}(0)$ 
in \eqref{D-min} unchanged.
Therefore the r.h.s. of \eqref{D-min} can be made smaller than any 
$\D > 0$, by choosing $\l$
small enough for large $\beta$ such that $\rho^{P}(\beta,-\D)$ gets small. 
Hence, we recover the result of 
Buffet, de Smedt and Pul\'e \cite{buffet:1983b}
about the stability of Bose-Einstein 
condensation in the weakly interacting Bose gases with the Van der Waals 
scaled potentials and a non-zero one-particle spectral gap. 

If $v(0)$ tends to infinity, 
then the condition \eqref{D-min} can not be satisfied for any finite gap. 
In this case our estimate becomes a triviality.  
This behaviour is compatible with the observation that 
for the hard core continuous Bose gas 
in a scaled external field 
there is no condensation 
with a macroscopic occupation of any level of the one-particle Hamiltonian 
\cite{macaonghusa:1987}.

Finally we remark that our results are for continuous homogeneous systems. 
The only assumptions we make are the gap in the one-particle excitations 
spectrum (\ref{T-Delta}) and the superstability conditions on the 
pair-potential $v$ (conditions \ab). 
Various other interesting exact results on Bose condensation are known,
\eg for Bose systems with a family of Van der Waals potentials 
\cite{buffet:1983b,desmedt:1987}, for models with truncated 
interactions \cite{dorlas:1991,dorlas:1993,zagrebnov:2001}, or 
for Bose lattice models with hard core interaction and at 
half-filling \cite{kennedy:1988}.
Only recently, a proof of BEC is found for the trapped interacting 
gases \cite{lieb:2002}, 
\ie for inhomogeneous systems, in the so-called Gross-Pitaevskii limit for
particle interactions.

\subsection*{Acknowledgments}
It is a pleasure to thank Geoffrey Sewell, Joe Pul\'e and Tony Dorlas for 
interesting comments. 
We are also thankful to one of the referees for a useful 
remark leading to some simplification of the presentation.
J.L. acknowledges financial support from  K.U.Leuven grant 
FLOF-10408, and V.A.Z. acknowledges ITF K.U.Leuven for hospitality.

\bibliographystyle{/loc_home/lauwers/artikels/myunsrt}
\bibliography{/loc_home/lauwers/artikels/biblio}

\end{document}